\documentclass[aps,prl,twocolumn,groupedaddress,amsmath,amssymb]{revtex4}
\usepackage{epsfig}
\usepackage{color}
\usepackage{subfigure}
\begin{document}

\newcommand{\bra}[1]{\langle #1|}
\newcommand{\ket}[1]{|#1\rangle}
\newcommand{\braket}[2]{\langle #1|#2\rangle}
\newcommand{\im}{\textrm{Im}\ }
\newcommand{\ff}[2]{\frac{1}{#1-#2}}
\newcommand{\qh}{\phi_{\mathrm{qh}}}
\newcommand{\el}{\phi_{\mathrm{el}}}
\newcommand{\qhv}{\sigma e^{i \varphi/2}}
\newcommand{\elv}{\psi \, e^{i \varphi}}
\newcommand{\be}{\begin{equation}}
\newcommand{\ee}{\end{equation}}
\newcommand{\fl}{{\Delta N_\phi}}
\newcommand{\spinor}[2]{\begin{pmatrix} #1 \\ #2 \end{pmatrix}}
\newcommand{\zb}{\overline{z}}
\newcommand{\up}{\uparrow}
\newcommand{\dn}{\downarrow}
\newcommand{\nn}{\nonumber}

%\preprint{}

\title{Spin texture readout of a Moore-Read topological quantum register}


\author{J.C. Romers}
\email{j.c.romers@uva.nl}
%\thanks{}
\author{K. Schoutens}
\affiliation{Institute for Theoretical Physics, University of Amsterdam,
Science Park 904, P.O.Box 94485, 1090 GL Amsterdam, The Netherlands}


\date{\today}

\begin{abstract}
We study the composite Charged Spin Texture (CST) over the Moore-Read quantum Hall state that arises when a collection of elementary CSTs are moved to the same location. Following an algebraic approach based on the characteristic pair correlations of the Moore-Read state, we find that the resulting CST is set by the fusion sector of the underlying non-Abelian quasiparticles. This phenomenon provides a novel way to read out the quantum register of a non-Abelian topologically ordered phase.
\end{abstract}

\pacs{}
%\keywords{}

\maketitle
\paragraph*{Introduction}
The observation of a fractional quantum Hall (fqH) effect at filling fraction $\nu = 5/2$ has led to a most interesting sequence of ideas and developments. The initial consensus, based on tilted field experiments \cite{PhysRevLett.61.997}, that the 5/2 state is not spin-polarized was reversed after numerical studies by Morf \cite{PhysRevLett.80.1505}. This lent support to the identification of the state underlying the 5/2 fqH effect with the Moore-Read (MR) (or Pfaffian) state first proposed in \cite{Moore1991}. Experimentally probing the 5/2 state is exceedingly difficult and, more than twenty years after the first observation, the issues of its spin-polarization and of the relevance of the MR state (or one of its close relatives such as the anti-Pfaffian state \cite{PhysRevLett.99.236806,PhysRevLett.99.236807}) have not been fully settled. Recent experimental evidence lends some support to the identification of the 5/2 state with the MR or anti-Pfaffian state and a growing body of numerical work in fact supports it. 

In two recent numerical studies the effect of the electron spin on the states near $\nu=5/2$ was explicitly taken into consideration \cite{PhysRevB.79.115322,PhysRevLett.104.086801}. The numerical findings of \cite{PhysRevLett.104.086801} are consistent with a physical picture where the groundstate at $\nu=5/2$ is a spin-polarized MR state, but where elementary excitations involve overturned spins. This picture is analogous to that for the integer quantum Hall (iqH) state at $\nu=1$, where the lowest energy excitations are charged spin textures (CST) known as \mbox{(anti-)}skyrmions \cite{PhysRevLett.75.4290}. These iqH skyrmions carry unit electric charge as well as unit topological charge, represented by the winding number (Pontryagin index) of the spin vector. In \cite{PhysRevLett.104.086801} the authors concluded that the CSTs over the MR state appearing at one excess flux can be of two types, depending on the interactions: the two underlying quasiholes can bind together and a skyrmion is formed, or they stay separate and form two separate CSTs.

In our recent paper \cite{1367-2630-13-4-045013}, we studied the charged spin textures associated with the elementary (charge $e^*=e/4$) 
excitations over the MR state by analytic means. We found that, while the CST associated with the charge $e/2$ excitations (akin to the Laughlin quasiparticles) are similar to the iqH skyrmions, those coming with the $e/4$ charges are a particular incarnation of `half-skyrmions'. Building on the fact that in the MR state the electrons are weakly bound into effective spin-1 bosons, we identified the $e/4$ CST with what are known as polar core vortices. In the core of such vortices the spin-1 bosons form the polar state (which is the mean field groundstate for $S=1$ bosons in the presence of antiferromagnetic interactions, with vanishing expectation of the spin vector), while away from the core the spins are fully polarized.

In the absence of spin, $e/4$ excitations over the MR state are known to behave as non-Abelions: the presence of $n$ such excitations leads to a total of $2^{n/2-1}$ topologically different states, all degenerate as long as the excitations are sufficiently separated. In the formalism where the MR wavefunctions are obtained as conformal blocks in a conformal field theory (CFT), this degeneracy can be traced to fact that the $e/4$ quasiparticles carry a $\sigma$ field, with Ising fusion rules $\sigma \times \sigma = 1 + \psi$. The topological distinction among the $2^{n/2-1}$ internal states of a collection of $n$ quasiholes over the MR  state makes them ideally suited to act as a (topologically protected) quantum register. Quantum gates can be implemented through particle braiding and readout is in principle possible by bringing the quasiparticles to one location and then probing the 1-particle density profile of the resulting composite excitation.

In this Letter we consider the same process of readout through fusion, but now in the presence of the spin degree of freedom. We thus analyze the thought-experiment where a collection of $n$ elementary charge $e/4$ CST are brought to the same location and fuse into a composite excitation of charge $ne/4$.  Focusing on the spin texture carried by the fused, composite excitation, we will establish a direct connection between the winding numbers characterizing the texture and the fusion sector of the underlying non-Abelions. This in principle enables the readout of a MR quantum register through the detection of a characteristic spin texture. 

There exist fermionic and bosonic versions of fqH wave functions; they differ by an overall Vandermonde factor.  Throughout this work, we focus on the bosonic analogue of the MR wavefunction. It has filling fraction $\nu=1$ and its elementary quasiholes come with charge $e^*=e/2$.

\paragraph*{Fusion of multiple MR quasiholes}
Excess flux $\fl$ over a MR state leads to $n=2\fl$ quasiholes. The fusion product of the corresponding $\sigma$ fields can result in 
$F$  $\psi$-quanta, with $0 \leq F \leq \fl$. These $\psi$ quanta will be absorbed by $F$ unpaired particles in the MR qH condensate. The number $F$ provides information about the topological state of the system: a determination of $F$ is equivalent to partially reading out the quantum register spanned by the fusion space of the quasiholes. If we now bring all quasiholes to the origin, the lowest surviving field 
product in the sector with $F$ $\psi$'s is $: \psi(0) \partial \psi(0) \ldots \partial^{F-1} \psi(0):$, where the colons denote normal ordering. 
This leads to the following ``big hole'' wave function, expressed in the form of a CFT correlator
\begin{widetext}
\be
\label{eq.cfthole}
\Psi_{\text{big hole}}(z_1 , \dots , z_N; \Delta N_\phi, F) = \prod_{i<j}(z_i-z_j) \prod_k z_k^\fl \langle : \psi(0) \partial \psi (0) \dots \partial^{F-1}\psi(0): \psi(z_1) \dots \psi(z_N) \rangle_{\rm CFT} .
\ee
\end{widetext}

In \cite{PhysRevB.54.16864} Read and Rezayi presented a general analysis of wavefunctions for quasiholes over the MR state, which they characterized by the same quantity $F$, the number of unpaired particles, and a set of integers $\{m_1,\ldots,m_F\}$. If these $m_k$ are chosen such that the highest obtainable value of the angular momentum $L_z$ is reached and all the quasihole coordinates are sent to the origin, their quasihole wavefunctions reduce to 
\begin{multline}\label{wf.f}
\prod_{i<j} (z_i-z_j) \prod_k z_k^\fl \sum_{\sigma \in S_N} \mathrm{sgn} \sigma z_{\sigma(1)}^{-1} \dots z_{\sigma(F)}^{-F} \\
\times \frac{1}{z_{\sigma(F+1)} - z_{\sigma(F+2)}} \dots \frac{1}{z_{\sigma(N-1)} - z_{\sigma(N)}} .
\end{multline}
It is easily seen that this expression is equivalent to Eq.~(\ref{eq.cfthole}).
%  In \cite{} Read and Rezayi introduce the same quantity $F$ as `the number of unpaired particles' and they provide the following 
%  general expression for the $n=2\fl$ quasihole states in the presence of $\fl$ extra fluxes
%  \begin{widetext}
%  \begin{eqnarray*}
%  \Psi_{m_1,\dots, m_F}(z_1,\dots ,z_N;w_1,\dots ,w_{2\fl}) = \frac{1}{2^{(N-F)/2}(N-F)/2!}\sum_{\sigma \in S_N}
%  \mathrm{sgn}\sigma \prod_{k=1}^F z_{\sigma(k)}^{m_k} \prod_{l=1}^{(N-F)/2} \\
%  \frac{\Phi(z_{\sigma(F+2l-1)},z_{\sigma(F+2l)};w_1, \dots , w_{2\fl} )}{(z_{\sigma(F+2l-1)}-z_{\sigma(F+2l)} )} \prod_{i<j}(z_i-z_j),
%  \end{eqnarray*}
%  \end{widetext}
%  where
%  \begin{eqnarray*}
%  \lefteqn{\Phi(z_1,z_2;w_1,\dots ,w_{2\fl} ) =} 
%  \\ &&
%  {1 \over \fl !^2} \sum_{\tau \in S_{2\fl}} \prod_{r=1}^{\fl} (z_1-w_{\tau (2r-1)}) (z_2 - w_{\tau(2r)}).
%  \end{eqnarray*}
% We recognize the equivalence of eq. (\ref{eq.cfthole}) and a special case of this expression, namely the one where the set ${m_k}$ 
% has been chosen such that the highest obtainable value of $L_z$ is reached and all the quasihole coordinates are sent to the origin. 
% This corresponds to the expression
% \begin{eqnarray*}\label{wf.f}
% \prod_{i<j} (z_i-z_j) \prod_k z_k^\fl \sum_{\sigma \in S_N} \mathrm{sgn} \sigma z_{\sigma(1)}^{-1} \dots z_{\sigma(F)}^{-F} \\
% \times
% \frac{1}{z_{\sigma(F+1)} - z_{\sigma(F+2)}} \dots \frac{1}{z_{\sigma(N-1)} - z_{\sigma(N)}},
% \end{eqnarray*}
% which is easily seen to be equivalent to eq. (\ref{eq.cfthole}).

In preparing for the introduction of the spin degree of freedom, we wish to rewrite the `big hole' wavefunction Eq.~(\ref{eq.cfthole}) in yet another form. The wave function for the MR state without quasiholes
\be
\label{eq.mrgs}
\Psi_{\rm{MR}} (z_1,\dots,z_N) = \prod_{i<j}(z_i-z_j) \text{Pf} \frac{1}{z_k - z_l},
\ee
is obtained by using the fqH-CFT connection with electron operator $\el (z) = \elv (z)$ in the $I\!sing \otimes U(1)$ CFT. It is the densest zero-energy eigenstate of the pairing Hamiltonian
\be
\label{eq.deltadelta}
H_{\rm pair} = \sum_{i<j<k}\delta(z_i-z_j) \delta(z_j-z_k),
\ee
which is saying that the wave function should not vanish when two coordinates are put equal whereas it should be identically zero when a third is set to the same value.

Eq.~\eqref{eq.mrgs} can be cast in another form that is especially suited for our purposes. We start by dividing the coordinates into two groups
\[
I = \{ z_1,\dots , z_{N/2} \}, \;\; I\! I = \{ z_{N/2+1},\dots , z_{N} \},
\]
and write down the following expression
\be 
\label{eq.mrgs-gp}
\Psi_{\rm MR} (z_1 , \dots , z_N) =
\begin{array}[t]{c} {\rm Symm} \\ {\small I,I\! I} \end{array} 
   \left[ \Psi^L_{I} \Psi^L_{I\! I} \right]\ ,
\ee
with a bosonic Laughlin wave function at $\nu=1/2$ for each group separately, 
$ 
\psi^L_{I,I\!I} (\{z_j\})=
\prod_{i<j \in I,I\!I} (z_i-z_j)^2 .
$
The symmetrization step in \eqref{eq.mrgs-gp} is performed over all possible partitions of the particles into the groups $I$ and $I \! I$. The equivalence of \eqref{eq.mrgs} and \eqref{eq.mrgs-gp} can be understood by arguing that both are densest zero-energy eigenstates of the Hamiltonian \eqref{eq.deltadelta}, that this densest zero-energy eigenstate is unique and that therefore the two wave functions are equal. Originally however \cite{Cappelli2001499}, the equivalence was found using a bosonization of the $I\! sing \otimes I\! sing$ CFT; we explain how this works because we wish to extend the result to the wavefunction \eqref{eq.cfthole} describing the fusion product of a collection of quasiholes.

The bosonization of the $I\! sing \otimes I\! sing$ CFT establishes a mapping between the field content of two independent copies of the $c=1/2$ $I\!sing$ CFT and a $c=1$ compactified boson, as shown below.
\[
\begin{array}{r c l c r c l}
I\! sing  & \otimes  & I\! sing & \leftrightarrow & \multicolumn{3}{c}{c=1 \text{ boson}}   \\
\hline
1 & \otimes &  1 & & & 1 & \\
\psi & \otimes & 1 & & { 1 \over \sqrt{2} } ( e^{i \varphi} & + & e^{- i \varphi} ) \\
1 &\otimes & \psi & & { 1 \over  i \sqrt{2} } ( e^{i \varphi} & - & e^{- i \varphi} ) \\
\sigma & \otimes & \sigma & & { 1 \over \sqrt{2} } ( e^{i \varphi/2} & + & e^{- i \varphi/2} )
\end{array}
\]
Let us first check how this formulation gives the MR ground state wave function. Up to an overall Vandermonde factor, the relevant CFT correlator is the one containing an even number of $\psi$ fields at each of the electron coordinates. In the bosonic description
\be
\label{eq.bosgroups}
\langle  \left( e^{i \varphi} + e^{- i \varphi} \right) (z_1)  \dots  \left( e^{i \varphi} + e^{- i \varphi} \right) (z_N)  \rangle,
\ee
only terms in which the number of factors of $e^{i \varphi}$ and $e^{-i \varphi}$ are equal contribute. We therefore select $N/2$ coordinates $I=\{ z_1 , \dots , z_{N/2} \} $ for which we take the positive exponential and take the negative exponential for the other half $I \! I = \{z_{N/2+1},\dots , z_N   \} $. This gives a contribution
\[
\prod_{i<j \in I} (z_i - z_j) \prod_{k<l \in I\! I} (z_k - z_l) \prod_{m \in I, n \in I\! I} (z_m - z_n)^{-1}.
\]
Combining this with the overall Vandermonde factor and summing over all permutations of the particle coordinates, since the partition into groups $I$ and $I \! I$ was completely arbitrary, we obtain the wave function \eqref{eq.mrgs-gp}.

To extend the `two group' formula to the `big hole' wave function Eq.~\eqref{eq.cfthole}, we need to bosonize the normal ordered 
field product  $:\psi(0) \partial \psi(0) \dots \partial^{F-1}\psi(0):$, which we will do as follows.
Let us locate the $\psi$ fields at locations $w_i$ and pull out all derivatives. The desired expression is then precisely the regular part of
\be
\partial_{w_2} \partial^2_{w_3} \dots \partial^{F-1}_{w_F} (e^{i \varphi} + e^{- i \varphi})(w_1) \dots(e^{i \varphi} + e^{- i \varphi})(w_F)
\nonumber
\ee
that survives after sending all the $w_i$ to zero. Following a similar logic as in the evaluation of Eq.~\eqref{eq.bosgroups} we must determine which terms in the expansion of this product contribute. The OPE of a positive and a negative exponential has singular behaviour (which the derivatives are only going to make stronger) and we maintain only regular terms due to the normal ordering. Thus only two terms contribute: 
one where we choose all the positive exponentials and one where we choose all the negative ones. The contribution of the positive exponentials
\[
\partial_{w_2} \partial^2_{w_3} \dots \partial^{F-1}_{w_F} \prod_{i<j}(w_i-w_j) e^{i\sum_k \varphi(w_k)},
\]
has only one term that survives when all $w_i$ are sent to zero, namely the one where all the derivatives act on the polynomial part (more precisely, the $w_1^0 w_2^1 \dots w_F^{F-1}$ term in its expansion). The same reasoning applies to the negative exponentials and one is left with the particularly simple identification
\be
\label{eq.ident}
:\psi(0) \partial \psi(0) \dots \partial^{F-1}\psi(0): \leftrightarrow (e^{iF\varphi}+e^{-iF\varphi})(0).
\ee
With this result, we can rewrite the wavefunction Eq.~\eqref{eq.cfthole} in the following way using the `two group' construction. To maintain charge neutrality we put $\frac{N-F}{2}$ particles in group $I$ and $\frac{N+F}{2}$ particles in group $I \! I$. The resulting expression is
\begin{multline}
\begin{array}[t]{c} {\rm Symm} \\ {\small I,I\! I} \end{array} 
\left[ \prod_{i,j \in I} (z_i-z_j)^2 \prod_{i \in I} z_i^{\Delta N_\phi + F}   \right. \\
\left. \times \prod_{k,l \in I \! I} (z_k-z_l)^2 \prod_{k \in I\! I} z_k^{\Delta N_\phi - F} \right].
\label{eq.symqh}
\end{multline}
One can interpret this result in a different way.
Since the number of flux quanta required to fit the Laughlin wave function for both of the groups on the sphere is different, both groups will have different excess flux.
For $N_\phi = N-2+\Delta N_\phi$, the particles in group $I$ have $\Delta N_\phi^I = \Delta N _\phi +F$ and those in group $I \! I$ have $\Delta N_\phi^{I \! I} = \Delta N _\phi -F$. We shall now demonstrate that the number of excess flux quanta per group determines the shape of the possible spin textures.

\paragraph*{Non-Abelian spin textures}
Let us first recapitulate the construction of CSTs explained in our previous paper. It is well-known that for iqH states the wave function for a skyrmion factorizes as
\be
\Psi_{\text{Skyrmion}} = \Psi_{\text {B}}  \times \Psi_{\text{iqH}},
\ee
where $\Psi_{\text{iqH }}$ is the ground state wavefunction for filling $\nu=1$ and $\Psi_{\text{B}}$ is a wavefunction for spinful
bosons in two orbitals \cite{PhysRevLett.76.2153}. In second quantization it is given by
\be
\label{eq.sk}
\ket{0,\up_N} + \lambda \ket{\dn,\up_{N-1}}+ \dots+ \lambda^N \ket{\dn_N,0}.
\ee
The two single particle angular momentum states in this expression are due to one extra flux quantum being present. The skyrmion resulting from this wave function has Pontryagin index (or winding number) 1; higher topological charge textures can be built by adding extra flux quanta and repeating the procedure. For a texture with winding number $w$, one has to add $w$ flux quanta to the ground state and take the superposition
\be
\label{eq.sk-w}
\ket{0,\dots,\up_N} + \lambda \ket{\dn,\dots,\up_{N-1}}+ \dots+ \lambda^N \ket{\dn_N,\dots,0}.
\ee
The crucial insight in our previous work was that when one uses the formulation \eqref{eq.mrgs-gp} of the Moore-Read state, it is possible to create textures labeled by independent winding numbers $[w_I,w_{I\! I}]$, one for each group of particles.

The elementary CST (in the sense that it carries the elementary electric charge $e^*$) was identified with winding numbers $[1,0]$ whereas the skyrmion, carrying electric charge $2 e^*$, has winding numbers $[1,1]$. In the core of a CST$[1,0]$, the 
spins of the particles in group $I$ (winding number 1) are overturned, whereas those in group $I\! I$ (winding number 0) are untouched. This then explains that the net spin in the core is vanishing. Switching to a notation where we specify a mean-field spinor of an associated spin-1 BEC, we identify the CST$[1,0]$ with $\xi(z) \propto (z,1,0)$. In contrast, the skyrmion CST$[1,1]$ corresponds to $\xi(z) \propto (z^2, \sqrt{2}\;z,1)$.

\paragraph*{Fusion channel to spin texture locking}
Combining the results specified in the above we arrive at the following claim for wavefunctions in the minimal-energy subspace 
defined by $H_{\text{pair}}\psi=0$: {\it the spin texture that is associated with the fusion product of $2\fl$ elementary CST over the MR state, in the fusion channel with $F$ unpaired particles, is a composite CST with electric charge $2 \fl\, e^*$ and winding numbers $[\fl+F,\fl-F]$.}

In general, the fusion channel label $F$ satisfies $(-1)^F = (-1)^N$. Thus, the simplest situation for $N$ odd is $F=1$ and $\fl=1$, leading to a CST labeled as $[2,0]$. % This shows that for odd $N$ the fully polarized state is not to be expected.

\begin{figure}[h!]
\begin{center}
\subfigure[{ CST[4,4]}]{\label{fig.44}
\includegraphics[width= .45\columnwidth]{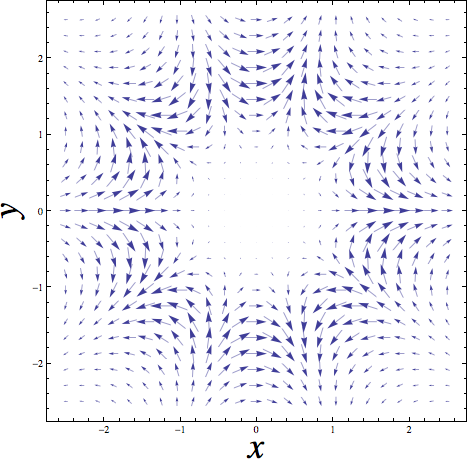}
\hspace*{2mm}
%\raisebox{0.8cm}
{\includegraphics[width= .45\columnwidth]{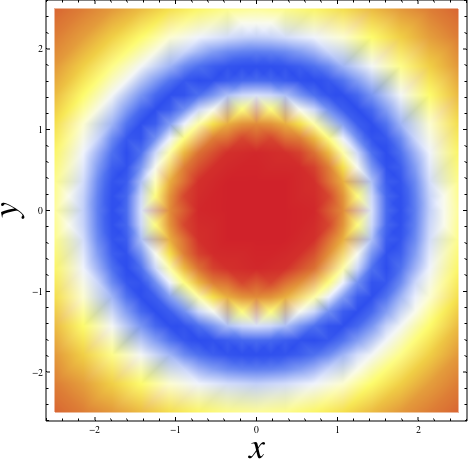}}
}
\subfigure[{ CST[6,2]}]{\label{fig.62}
\includegraphics[width= .45\columnwidth]{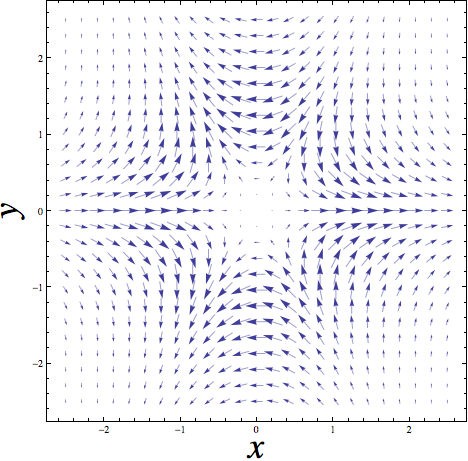}
\hspace*{2mm}
%\raisebox{0.8cm}
{\includegraphics[width= .45\columnwidth]{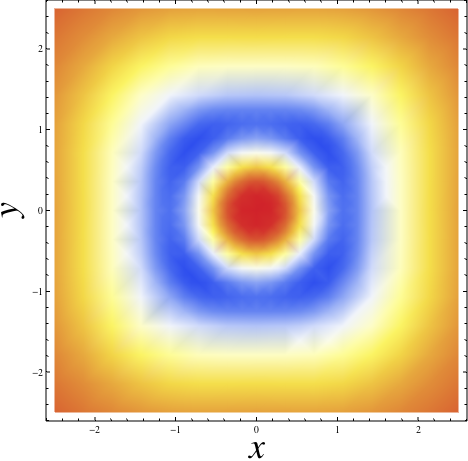}}
}
\subfigure[{ CST[8,0]}]{\label{fig.80}
\includegraphics[width= .45\columnwidth]{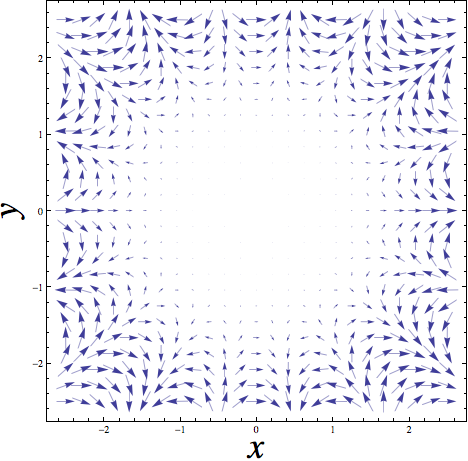}
\hspace*{2mm}
%\raisebox{0.8cm}
{\includegraphics[width= .45\columnwidth]{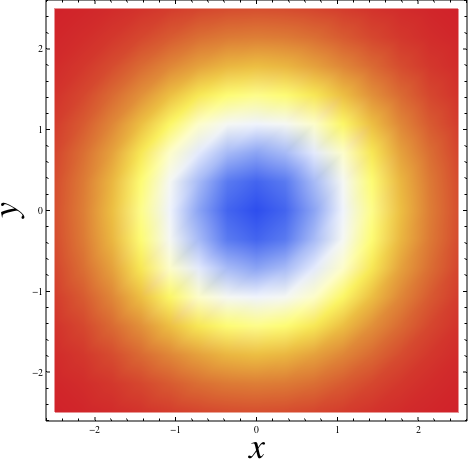}}
}
\caption{Stereographic projection of the $(S_x,S_y)$ vector field (left) and the length of the spin vector $S^2$ (right) for spin textures CST[4,4], CST[6,2] and CST[8,0]. All CSTs have $\lambda_1=\lambda_2 =0.5$ and $N=8$. In the $S^2$ plots, warmer colors correspond to a longer length of the spin vector. The normalization is such that $S^2=1$ far away from the core of the CST.} 
\label{fig:N8textures}
\end{center}
\end{figure}

In Figures \ref{fig.44}--\subref{fig.80} we display the the spin textures for $N=8$ particles in the presence of $\fl=4$ excess flux quanta. Possible sectors are $F=0$ with CST$[4,4]$, $F=2$ with CST$[6,2]$ and $F=4$ leading to CST$[8,0]$. %OPTIONAL: extra sentence explaining how spin textures are reproduced using CP2 spinor formulation.

\paragraph*{Spin texture readout}
In the fully polarized MR state the fingerprint left by the unpaired fermions takes the form of an altered density profile \cite{PhysRevB.80.115121} and energy difference \cite{PhysRevLett.107.036803} with respect to the state without unpaired fermions. We find that, within the framework we used, the textures formed by the electron spins are unique and in 1-1 correspondence with the number $F$ characterizing the fusion sector of the non-Abelian CST. This provides a novel method to determine the number of unpaired fermions and thus to read out the ($F$-number of the) topological quantum register.

In the present work, we have circumvented the task of minimizing the realistic (Coulomb and Zeeman) interaction energies by working within the zero energy subspace of the pairing Hamiltonian \eqref{eq.deltadelta}. The rationale for this is the well-established fact that in the spin-polarized case the correlations found in systems with realistic interactions agree with those enforced by the pairing condition $H_{\text{pair}}\psi=0$.
In addition, we have relied on the `two group' Ansatz and have assumed that the particles in each group $I$, $I\! I$ separately maximize their winding number. 

It remains to be confirmed that our results capture the essence of what happens when realistic (Coulomb and Zeeman) interactions are used instead of the pairing condition. A numerical approach is possible in principle but very challenging in practice. The spin degree of freedom makes the use of exact diagonalization highly nontrivial due to the size of the Hilbert spaces involved in the problem. A Monte Carlo study is hindered by the fact that no simple first-quantized expressions have been found yet for the CST wave functions.

\paragraph*{Acknowledgements}We thank Steve Simon for inspiring discussions. JCR and KJS would like to acknowledge the hospitality of the Institut Henri Poincar\'{e}. JCR is financially supported by FOM.

\bibliography{research}
\end{document}